\documentclass[a4paper, 12pt]{article}
\usepackage{graphicx}
\begin{document}

\begin{center}
{\Large Generation of polarization entangled photon pairs and violation of Bell's inequality using 
spontaneous four-wave mixing in fiber loop}
\end{center}

\begin{center}
Hiroki Takesue and Kyo Inoue
\end{center}

\begin{center}
NTT Basic Research Laboratories, NTT Corporation.\\
3-1 Morinosato-Wakamiya, Atsugi, Kanagawa, 243-0198, Japan.\\
E-mail: htakesue@will.brl.ntt.co.jp\\
\vspace{5mm}
August 5, 2004
\end{center}
\vspace{10mm}
\begin{center}
Abstract
\end{center}
We report the generation of polarization entangled photon pairs in the 1550-nm wavelength band 
using spontaneous four-wave mixing in a dispersion-shifted fiber loop. The use of the fiber-loop 
configuration made it possible to generate polarization entangled states very stably. With accidental 
coincidences subtracted, we obtained coincidence fringes with $>$90 \% visibilities, and observed 
a violation of Bell's inequality by seven standard deviations. 
We also confirmed the preservation of the quantum correlation between the photons even after they 
had been separated by 20 km of optical fiber.

\newpage
In recent years, the generation of entangled photon pairs has been an important topic, because it is 
regarded as a key technology for realizing quantum communication, including quantum 
cryptography \cite{bennett1}, quantum dense coding \cite{bennett2}, and quantum teleportation 
\cite{bennett3}. 
There are several kinds of entangled photon pairs based on different variables, of which 
polarization entanglement has been studied intensively \cite{aspect}-\cite{li} because it is 
relatively easy to deal with in experiments. 
Most reported experiments have been undertaken in the 700-nm wavelength band using parametric 
down-conversion (PDC) with type-II phase matching, which was first demonstrated by Kwiat et al. 
\cite{kwiat1}. 
They also reported a polarization entanglement experiment that used PDC with type-I phase 
matching in a two-crystal geometry also in the 700-nm wavelength band \cite{kwiat2}, 
\cite{white}. 
In addition, several important experiments have been reported in relation to polarization 
entanglement, such as quantum cryptography \cite{jen} and entanglement formation using a 
quantum dot single photon source and post-selection \cite{fattal}. 
However, to realize ``quantum communication systems" over optical fiber networks, it is very 
important to generate entangled photon pairs in the 1550-nm band, where conventional silica fiber 
has its minimum loss.
In this wavelength band, Yoshizawa et al. reported the generation of polarization entangled photon 
pairs using PDC in two periodically poled lithium niobate (PPLN) waveguides and post-selection 
\cite{yoshi}. 
Recently, Li et al. demonstrated the generation of polarization entangled photon pairs using 
spontaneous four-wave mixing (FWM) in a fiber Sagnac loop \cite{li}. They used time-delayed 
orthogonally-polarized pump pulses to generate two product states, and then removed the time 
distinguishability by passing the product states through a birefringent fiber. 
These methods required precise control circuits to stabilize the relative phase between two product 
states, and this made the systems complicated.  

In this Letter, we propose and demonstrate a method for generating polarization entangled photon 
pairs in the 1550-nm band. Our method uses spontaneous FWM in a fiber loop formed with a 
polarization beam splitter (PBS) and a dispersion shifted fiber (DSF), which is a single-mode 
optical fiber whose zero-dispersion wavelength is set at around 1550 nm. The use of a loop 
configuration made it possible to stabilize the relative phase between two product states without the 
need for a control circuit, and thus we were able to generate polarization entangled states stably. In 
addition, we constructed our experimental setup using only fiber connections, which also 
contributed to the stable operation. 
We confirmed the feasibility of our method experimentally and observed a clear violation of Bell's 
inequality by seven standard deviations. Moreover, we observed the preservation of a quantum 
correlation between signal and idler photons even after they had been separated by 20 km of optical 
fiber.

%\section{Principle}

%four-wave mixing
Spontaneous FWM is a third-order optical nonlinear process, in which two pump photons are 
annihilated and a signal-idler photon pair is generated \cite{fiorentino,inoue1}. 
Here, we consider a partially degenerate case, in which the optical frequencies of the two pump 
photons are the same. 
Then, the pump photon, $\omega_p$, signal photon $\omega_s$, and idler photon 
$\omega_i$ frequencies satisfy the following relationship.
\begin{equation}
2 \omega_p = \omega_s + \omega_i \label{freq}
\end{equation}
We use a single-mode fiber as a nonlinear medium.
A single-mode fiber exhibits slight birefringence caused by small deviations from cylindrical 
geometry or small fluctuations in material anisotropy. Moreover, in a long fiber, the principal axis 
randomly changes along the fiber due to external perturbations, such as microbending and local 
pressure. 
In such fibers, only signal and idler photons whose polarization states are the same as those of the 
pump photons are generated \cite{inoue2}. Therefore, for example, if we input pump photons with 
a horizontal polarization ($H$) into a fiber, we can obtain a signal-idler product state $|H\rangle_s 
|H\rangle_i$ through a spontaneous FWM process. 

The following phase matching condition must be satisfied for the effective generation of photon 
pairs. 
\begin{equation}
2 k_p = k_s + k_i \label{pm}
\end{equation} 
Here, $k_p$, $k_s$ and $k_i$ denote the wavenumbers of the pump, signal and idler photons, 
respectively. 
When the power dependent refractive index is negligible, we can achieve the phase-matching 
condition in the 1550-nm wavelength band by using a DSF and setting the pump wavelength at the 
zero-dispersion wavelength \cite{inoue3}. 
Thus, the use of the DSF makes it possible to generate the photon pairs effectively in the 1550-nm 
wavelength band. 

\begin{figure}[thb]
\vspace{0mm}
\begin{center}
\includegraphics[width=.9\linewidth]{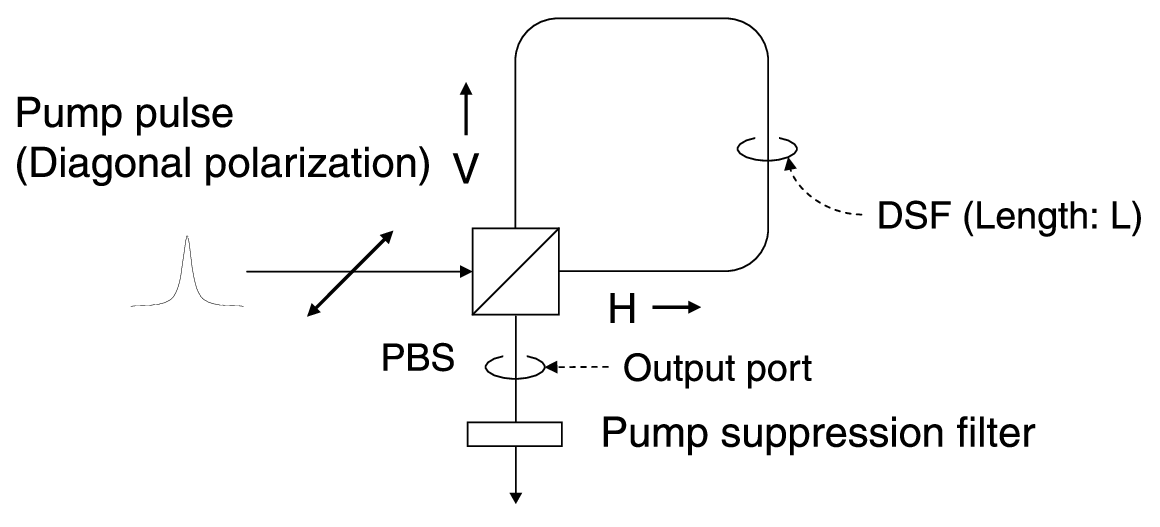}
\end{center}
\vspace{-6mm}
\caption{Schematic diagram of proposed method.}　\label{1}
\vspace{-3mm}
\end{figure}

%loop configuration
In order to generate a polarization entangled state, we place the DSF in a loop configuration with a 
PBS, as shown in Fig. 1. 
A pump pulse with a 45$^{\circ}$ or 135$^{\circ}$ linear polarization state is input into the fiber 
loop. 
The PBS divides the pump pulse into horizontal ($H$) and vertical ($V$) polarization components 
of equal power. 
The $H$ and $V$ components generate signal-idler photon pairs $|H\rangle_s |H\rangle_i$ and 
$|V\rangle_s |V\rangle_i$ while propagating in the loop in the counter-clockwise (CCW) and 
clockwise (CW) directions, respectively.  The photon pairs then move to the output port of the 
PBS (see Fig. 1). 
With an appropriate pump power, we can make the probability of generating two pairs 
simultaneously very low. As a result, a superposed state of the two product states $|H\rangle_s 
|H\rangle_i$ and $|V\rangle_s |V\rangle_i$ is output from the PBS. 
After eliminating the pump light using an optical filter, we can obtain a polarization entangled state. 
Note that this loop and the Sagnac loop in \cite{li} have different functions. The Sagnac loop in 
\cite{li} is employed solely to suppress the pump light, while our loop provides fundamental 
functions for entanglement generation, namely the path separation of the pump and the removal of 
path distinguishability between the two product states \cite{note}.  

%phase cancellation
In order to obtain a maximally entangled state, the relative phase between $|H\rangle_s 
|H\rangle_i$ and $|V\rangle_s |V\rangle_i$ should be 0 or $\pi$ at the loop output. 
As we explain below, this condition is automatically satisfied in our proposed method, owing to the 
fiber loop configuration. 
We assume that the signal and idler photon states at the generation point are expressed as $e^{i 
\phi_{sX}} |X\rangle_s$ and $e^{i \phi_{iX}} |X\rangle_i$, respectively. 
Here, $\phi_{sX}$ and $\phi_{iX}$ are the phases of the signal and idler photons with 
$X$ denoting $H$ or $V$ polarizations, which corresponds to the polarization states of the photons 
circulating in the CCW and CW directions, respectively. 
The signal and idler phases at the generation point, $\phi_{sX}$ and $\phi_{iX}$, satisfy the 
following relationship,  
\begin{equation}
\phi_{sX} + \phi_{iX}=2 \phi_{pX}, \label{phase} 
\end{equation}
where $\phi_{pX}$ denotes the phase of the pump photon at the generation point, which can be 
expressed as 
\begin{eqnarray}
\phi_{pH} &=& k_{pH} x, \label{pccw} \\
\phi_{pV} &=& k_{pV} (L-x)+\varphi.
\end{eqnarray}
Here, $k_{pX}$, $L$, and $x$ are the wavenumber of the pump, the loop length and the distance 
between the PBS and the generation point measured in the CCW direction in the loop, respectively. 
$\varphi$ is the initial phase difference between the $H$ and $V$ components of the pump photon 
at the loop input. 
The generated signal and idler photons propagate along the loop and appear at the output port. 
With $k_{sX}$ and $k_{iX}$ denoting the wavenumbers of the signal and idler photons, 
respectively, the photons generated in the CCW direction have the phases shown with the following 
equations at the output port. 
\begin{eqnarray}
\phi_{sH(out)} &=& k_{sH} (L-x) + \phi_{sH} \\
\phi_{iH(out)} &=& k_{iH} (L-x) + \phi_{iH}
\end{eqnarray}
Similarly, the phases of those generated in the CW direction are 
\begin{eqnarray}
\phi_{sV(out)} &=& k_{sV} x + \phi_{sV}, \\
\phi_{iV(out)} &=& k_{iV} x + \phi_{iV}. \label{icwout}
\end{eqnarray} 

The states of the signal photon and idler photons at the PBS output are expressed as
\begin{eqnarray}
|X\rangle_{s(out)} &=& e^{i \phi_{sX(out)}} |X \rangle_s, \\
|X\rangle_{i(out)} &=& e^{i \phi_{iX(out)}} |X \rangle_i. \label{stateout}
\end{eqnarray}
We assume that the power ratio between $H$ and $V$ photon pairs is $\alpha^2:\beta^2$, where 
$\alpha$ and $\beta$ are real and satisfy $\alpha^2+\beta^2=1$. We can control this ratio by 
adjusting the pump polarization. 
As a result, the entangled state obtained at the PBS output is expressed as
\begin{eqnarray}
|\Psi_{out}\rangle &=& \alpha |H \rangle_{s(out)} |H \rangle_{i(out)} + \beta |V 
\rangle_{s(out)} |V \rangle_{i(out)} \nonumber \\
&=& \exp\{i(\phi_{sH(out)}+\phi_{iH(out)} )\} \alpha |H\rangle_s |H\rangle_i \nonumber \\
& & + \exp\{i(\phi_{sV(out)}+\phi_{iV(out)})\} \beta |V\rangle_s  |V\rangle_i \nonumber \\
&=& \alpha |H\rangle_s |H\rangle_i+ e^{i \phi_r} \beta |V\rangle_s |V\rangle_i,  \label{ent} 
\end{eqnarray}
where the relative phase $\phi_r$ is expressed as
\begin{equation}
\phi_r = 2(k_{pV}-k_{pH}) L - \Delta k_V x + \Delta k_H (L-x) + 2 \varphi,  \label{relative}
\end{equation}
with the phase mismatching $\Delta k_X$ given by
\begin{equation}
\Delta k_X = 2k_{pX} - (k_{sX}+k_{iX}).
\end{equation}
We used Eqs. (\ref{phase}) - (\ref{stateout}) and eliminated the common phase term between the 
two product states in the transformation from the 2nd to the 3rd lines of Eq. (\ref{ent}). 
We assume that the phase matching condition Eq. (\ref{pm}) is satisfied for both polarization 
components, which is valid when the frequency dependence of the fiber birefringence is small. 
Then, the output state Eq. (\ref{ent}) is reduced to 
\begin{equation}
|\Psi_{out}\rangle = \alpha |H\rangle_s |H\rangle_i+ \beta \exp \left\{2 i \left(\varphi+ 
\frac{\omega_p \Delta n L}{c}\right) \right\}|V\rangle_s |V\rangle_i.  \label{ent2}
\end{equation}
Here, $\Delta n$ is the difference between the refractive indices of the $H$ and $V$ components.
Equation (\ref{ent2}) shows that the relative phase between the two product states does not depend 
on $x$, the point at which the photon pair is generated. 
This means that the relative phase is automatically adjusted and polarization entangled photon pairs 
are obtained stably. 
In addition, we can obtain the maximally entangled states $|\Phi^{\pm}\rangle = (|H\rangle_s 
|H\rangle_i \pm |V\rangle_s |V\rangle_i)/\sqrt{2}$ by adjusting the pump polarization to achieve 
$\alpha=\beta=1/\sqrt{2}$ and $\phi_r=0, \pi$.

%\section{Experiment}
\vspace{0mm}
\begin{figure}[thb]
\vspace{-2mm}
\begin{center}
\includegraphics[width=\linewidth]{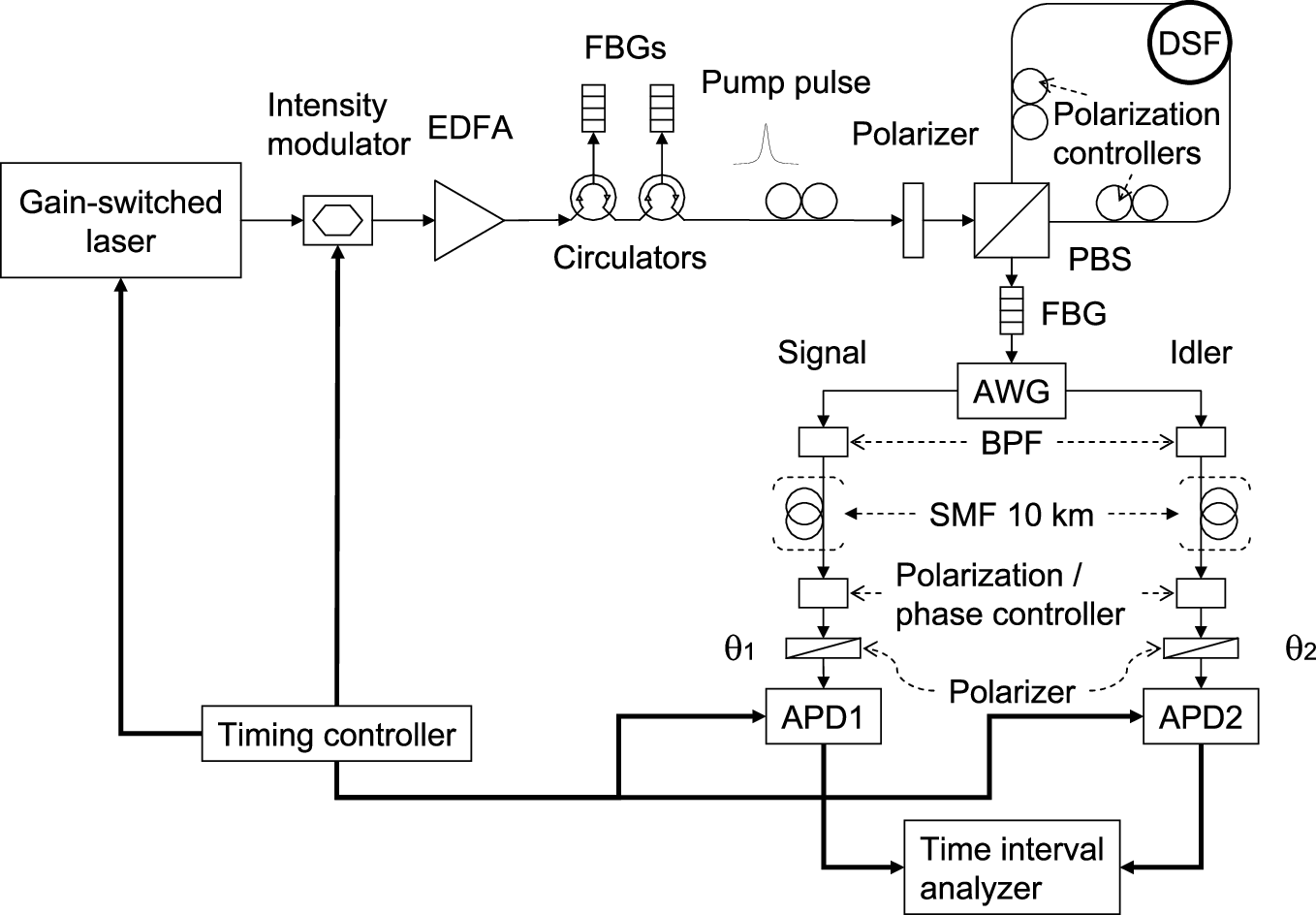}
\end{center}
\vspace{-7mm}
\caption{Experimental setup.}　\label{2}
\vspace{-3mm}
\end{figure}

We undertook experiments to confirm the effectiveness of our method using the setup shown in Fig. 
2. 
Pump pulses with a 20-ps width and a 2-GHz repetition frequency were generated using a 
gain-switched semiconductor laser. The center wavelength of the pulses was set at 1551 nm, 
namely, close to the zero-dispersion wavelength of the DSF used in the experiment. 
The pump pulses were input into an optical intensity modulator, which reduced the repetition 
frequency to 100 MHz, and were amplified using an erbium-doped fiber amplifier (EDFA). The 
amplified spontaneous emission (ASE) noise in the EDFA output was eliminated using narrowband 
fiber-Bragg gratings (FBG) and optical circulators cascaded in pairs. 
The state of polarization of the pulses was adjusted to 45$^{\circ}$ linear polarization (i.e. 
$\alpha=\beta=1/\sqrt{2}$ and $\varphi=0$ in Eq. (\ref{ent})) using a polarization controller (a 
quarter-wave plate (QWP) combined with a half-wave plate (HWP)) and a polarizer. 
Then the pulses were launched into the loop, which consisted of a PBS, a 2.5-km DSF and two 
polarization controllers. The peak power of the pump pulse was 42 mW at the input of the DSF.  
We adjusted the polarization controllers in the loop so that the generated photon pairs were properly 
output from the loop. 

The output photons from the loop were input into narrowband FBGs to suppress the pump photons, 
and were then launched into an arrayed waveguide grating (AWG) with a 50-GHz channel spacing 
to separate the signal and idler photons. 
AWG output channels with peak wavelengths of 1552.7 and 1549.4 nm were used for the signal 
and idler, respectively. 
The output photons from the AWG were filtered using dielectric optical bandpass filters (BPF) to 
further suppress the pump photons. With the FBGs, AWG and BPF, the pump photons were 
suppressed by $>$125 dB relative to the signal and idler photons. 
The 3-dB bandwidths of the signal and idler were
both approximately 25 GHz. 
This value was limited mainly by the width of the AWG passband.

The output photons from each BPF were then input into a polarization / phase controller, which 
consisted of a QWP, an HWP and a Babinet-Soleil compensator. 
Here, the polarization states of the signal and idler photons were adjusted so that the two photons 
experienced the same polarization change after they were separated by the AWG. 
Each photon was then input into a rotatable polarizer and detected with an avalanche photodiode 
(APD) operated in a gated Geiger mode. 
The electric signals from the APDs were input into a time interval analyzer for coincidence 
measurements. The quantum efficiency, gate width and repetition frequency of the APDs were 10 
\%, 2.5 ns and 1 MHz, respectively. The dark count probabilities per gate were $2.5 \times 
10^{-5}$ for APD1 (for signal) and $4.0 \times 10^{-5}$ for APD2 (for idler). 
The photon pair production ratio per pump photon was $\sim 6 \times 10^{-10}$, and the average 
count rates for the signal and idler were 490 and 380 cps, respectively. We would like to emphasize 
that the whole system was constructed with fiber connections, and thus quite stable.

We rotated the angle of the polarizer for the signal $\theta_1$ while fixing that for the idler 
$\theta_2$ at 0$^{\circ}$, and measured the coincidence rate $C(\theta_1,\theta_2)$. The result 
is shown by the circles in Fig. 3. The horizontal axis shows the $\theta_1$ value, and the vertical 
axis shows the coincidence rate per signal count. Accidental coincidence counts were subtracted.
A coincidence fringe with a visibility of 94.2 \% was observed. 
In order to confirm the quantum correlation, we undertook the same experiment for 
$\theta_2=45^{\circ}$. As shown by the squares in Fig. 3, we obtained a coincidence fringe with 
91.2\%-visibility. 
From these results, we can conclude that our method successfully generated polarization entangled 
photon pairs. 
The phase relationship between the two coincidence curves indicated that the obtained state was 
$|\Phi^{+}\rangle = (|H\rangle_s |H\rangle_i + |V\rangle_s |V\rangle_i)/\sqrt{2}$. The high 
visibility indicated that the birefringence of the DSF $\Delta n$ was so small that we could make 
the relative phase in Eq. (\ref{ent2}) nearly zero. 
When the accidental coincidence counts were included, the visibilities were 77.3 and 77.0 \% for 
$\theta_2=0$ and $45^\circ$, respectively. 
There were several reasons for these accidental coincidence counts.
They occurred when two pairs were generated by a pump pulse. In addition, noise photons were 
generated in the DSF through nonlinear processes such as Raman scattering \cite{inoue1} and self 
phase modulation of the pump pulse. These noise photons may also have caused the accidental 
coincidence. 

The maximum coincidence count rate was approximately 1.8 cps for 
$\theta_1=\theta_2=0^{\circ}$. This relatively low coincidence rate was mainly due to the large 
loss caused by the FBGs, AWG, BPFs and polarization controllers and the small quantum efficiency 
of the APDs in the 1550-nm band.

\begin{figure}[thb]
\vspace{3mm}
\begin{center}
\includegraphics[width=.7\linewidth]{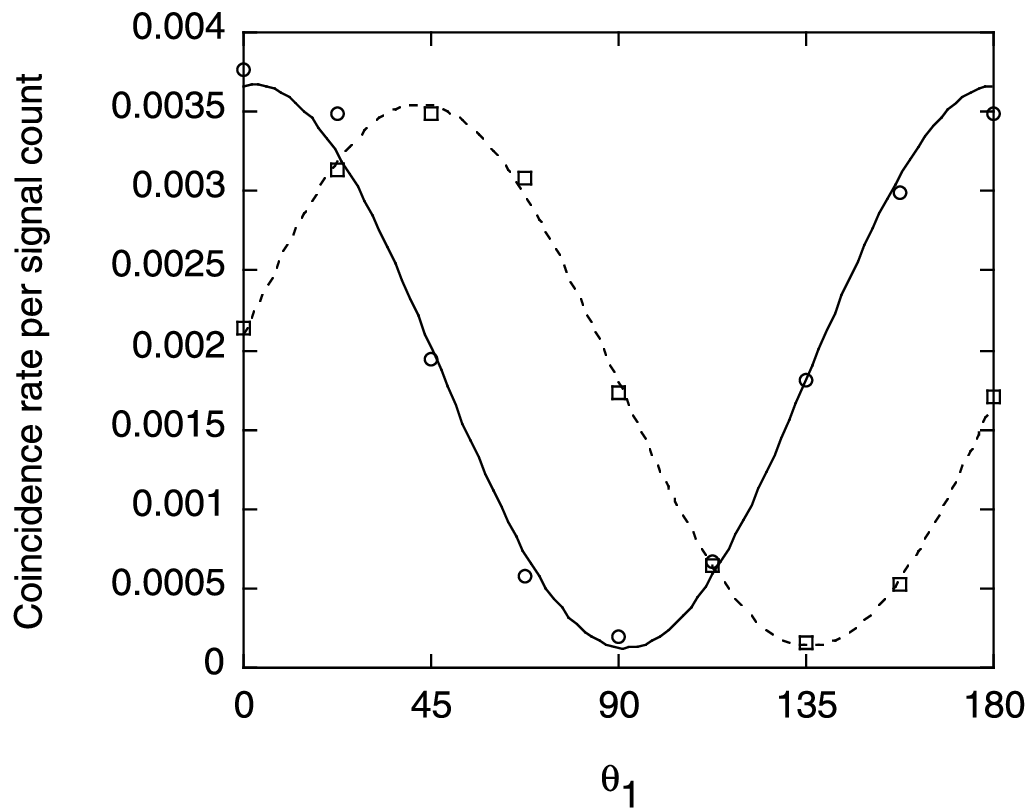}
\end{center}
\vspace{-6mm}
\caption{Coincidence rate per signal count as a function of $\theta_1$. Circles: 
$\theta_2=0^{\circ}$, squares: $\theta_2=45^{\circ}$. }　\label{3}
\vspace{-3mm}
\end{figure}
\begin{figure}[h]
\vspace{0mm}
\begin{center}
\includegraphics[width=.7\linewidth]{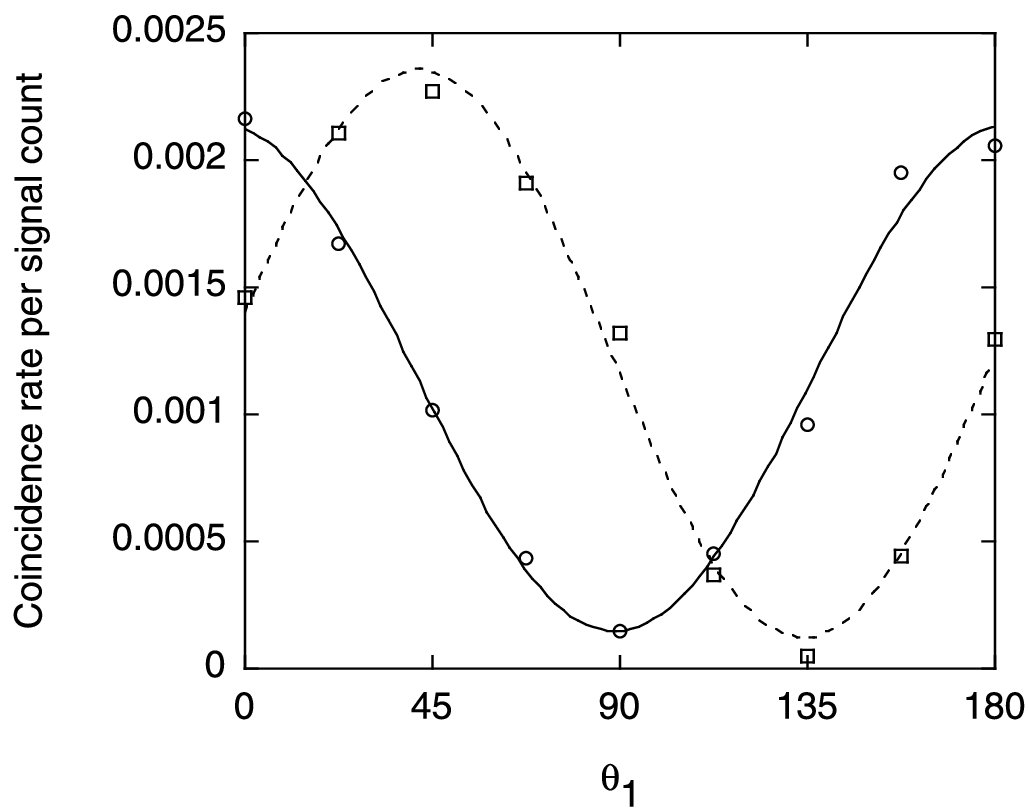}
\end{center}
\vspace{-6mm}
\caption{Coincidence rate per signal count as a function of $\theta_1$ after the signal and idler 
photons had been separated by 20 km of optical fiber. Circles: $\theta_2=0^{\circ}$, squares: 
$\theta_2=45^{\circ}$. }　\label{4}
\vspace{-3mm}
\end{figure}

We then performed a test of Bell's inequality. According to the inequality of Clauser, Horne, 
Shimony, and Holt (CHSH) \cite{CHSH}, $|S|$, defined below, should be less than 2 for any local 
realistic theory.
\begin{equation}
S = E(\theta_1, \theta_2) + E(\theta'_1, \theta_2)+ E(\theta_1, \theta'_2)- E(\theta'_1, \theta'_2)
\end{equation}
Here, $E(\theta_1, \theta_2)$ is given by
\begin{equation}
E(\theta_1, \theta_2)= \frac{C(\theta_1, \theta_2)+ C(\theta^{\perp}_1, \theta^{\perp}_2)- 
C(\theta_1, \theta^{\perp}_2)- C(\theta^{\perp}_1, \theta_2)}{ C(\theta_1, \theta_2)+ 
C(\theta^{\perp}_1, \theta^{\perp}_2)+ C(\theta_1, \theta^{\perp}_2)+ C(\theta^{\perp}_1, 
\theta_2)}.
\end{equation}
We measured the $S$ value for the settings $\theta_1=-22.5^{\circ}$, 
$\theta^{\perp}_1=67.5^{\circ}$, $\theta'_1=22.5^{\circ}$, $\theta'^{\perp}_1=112.5$; and 
$\theta_2=-45^{\circ}$, $\theta^{\perp}_2=45^{\circ}$, $\theta'_2=0^{\circ}$, 
$\theta'^{\perp}_2=90^{\circ}$. We undertook five runs of the $S$ value measurement, and 
obtained $S=2.65 \pm 0.09$ when we subtracted accidental coincidences.
Thus, we observed a violation of the CHSH inequality by seven standard deviations. When 
accidental coincidences were included, the $S$ value was $2.06 \pm 0.08$, which means that we 
still observed the violation by a 0.75 standard deviation.
In this experiment, a set of the $S$ values measurement with 16 pairs of polarizer angles took as 
long as about 50 minutes because of the low coincidence rate.
We would like to emphasize that we were able to observe a violation of Bell's inequality despite the 
long measurement time. 
This indicates that our method generated polarization entangled photons stably for as long as 
several hours without any major relative phase fluctuation, owing to the fiber-loop configuration.

%fiber transmission
Finally, we inserted a 10-km single-mode fiber after the BPF in each path (as shown in Fig. 2), and 
again measured the coincidence rate as a function of $\theta_1$ for $\theta_2=0^\circ$ and 
$45^\circ$. The result is shown in Fig. 4, from which accidental coincidences were subtracted. The 
obtained visibilities were 87.2 \% for $\theta_2=0^\circ$ and 90.1 \% for 
$\theta_2=45^\circ$ \cite{note2}. Thus, we were able to confirm that the strong quantum 
correlation between the signal and idler photons was preserved even after they had been separated 
by 20 km of optical fiber. 
This result reveals the possibility that polarization entangled photons generated using our method 
may be useful for long-distance quantum teleportation, in which a sender and a receiver need to 
share entangled particles.

%\section{Conclusion}

In summary, we proposed and demonstrated the generation of polarization entangled photons in the 
1550-nm wavelength band using spontaneous FWM in a DSF loop. 
The fiber-loop configuration enabled us to stabilize the relative phase making the stable generation 
of photon pairs possible. In addition, the whole setup was constructed with fiber connections, 
which contributed to the stable operation. 
With accidental coincidences subtracted, we obtained coincidence fringes with visibilities of more 
than 90 \% and observed the violation of Bell's inequality by seven standard deviations. We also 
confirmed that the strong quantum correlation between the photons was preserved after they had 
been separated by 20 km of optical fiber. 
We expect that our method will be a key technology for future quantum information systems 
implemented over optical fiber networks. 

%\section{Acknowledgement}

We thank Dr. K. Shimizu for helpful discussions. This work was supported in part by National 
Institute of Information and Communications Technology (NICT) of Japan.


\begin{thebibliography}{99}
\bibitem{note} In contrast, the method reported in \cite{li} is based on the time separation of the 
pump and the removal of the time distinguishability between two product states.
\bibitem{note2} The visibilities were 66.7 \% ($\theta_2=0^\circ $) and 70.6 \% 
($\theta_2=45^\circ$) when accidental coincidences were included.
\end{thebibliography}
\end{document}